# Absence of long-range *Ni/Mn* ordering in ferromagnetic La$_2$NiMnO$_6$ thin films

M.P. Singh, C. Grygiel, W.C. Sheets, Ph. Boullay, M. Hervieu,

W. Prellier,[*] B. Mercey, Ch. Simon, and B. Raveau

*Laboratoire CRISMAT, CNRS UMR 6508, ENSICAEN,*

*6 Bld. Maréchal Juin, F-14050 Caen, France*

(Dated: 07 June 2007)

Epitaxial La$_2$NiMnO$_6$ thin films have been grown on (001)-orientated SrTiO$_3$ using the pulsed laser deposition technique. The thin films samples are semiconducting and ferromagnetic with a Curie temperature close to 270 K, a coercive field of 920 Oe, and a saturation magnetization of 5 $\mu_B$ per formula unit. Transmission electron microscopy, conducted at room temperature, reveals a majority phase having "*I*-centered" structure with $a \approx c \approx a_{sub}\sqrt{2}$ and $b \approx 2a_{sub}$ along with minority phase domains having a "*P*-type" structure ($a_{sub}$ being the lattice parameter of the cubic perovskite structure). A discussion on the absence of *Ni/Mn* long-range ordering, in light of recent literature on the ordered double-perovskite La$_2$NiMnO$_6$ is presented.

The double perovskite $La_2NiMnO_6$, a magnetic semiconductor, has received considerable attention recently because the material demonstrates ferromagnetic order near room temperature with a Curie temperature ($T_C$) of 280 K.[1,2] The development of magnetic semiconductors exhibiting near room temperature ferromagnetism offers potential applications in next-generation spintronic devices, such as spin-based transistors and advanced magnetic memory storage elements. Moreover, large magnetic-field-induced changes in the resistivity and dielectric properties have been reported for single-phase samples of $La_2NiMnO_6$.[3] Materials that respond to multiple external stimuli, such as switching the dielectric constant with an applied magnetic field (magneto-dielectric effect), would enable novel device applications.

Samples of $La_2NiMnO_6$ have been studied extensively for their structural and physical properties. Disagreement exists on the basic structure of $La_2NiMnO_6$, including both the atomic *Ni/Mn* B-site sublattice order and cation oxidation state (i.e., $Ni^{2+}/Mn^{4+}$ or $Ni^{3+}/Mn^{3+}$). Early studies on bulk samples of $La_2NiMnO_6$ concentrated on verification of Goodenough-Kanamori's (GK) rules and synthesis of the first ferromagnetic insulator.[1,4] Instead, the $La_2NiMnO_6$ samples showed considerable electrical conduction and no sign of atomic ordering on the B-site sublattice. Thus, the authors concluded that ferromagnetic behavior observed in $La_2NiMnO_6$ samples resulted from $Mn^{3+}$-O-$Mn^{3+}$ superexchange interactions. Later, Blasse provided evidence based on magnetic susceptibility data that the preferred valency of $La_2NiMnO_6$ samples appeared to be $Ni^{2+}$ and $Mn^{4+}$ with some order on the B-site sublattice.[5] Since then, numerous reports, including magnetic, $^{55}Mn$ NMR, and X-ray absorption spectroscopy studies,[6-8] have provided further evidence for ordered $Mn^{4+}$-O-$Ni^{2+}$ superexchange interactions in $La_2NiMnO_6$, while two neutron diffraction studies disagree about the manganese and nickel oxidation states. Blasco et al. report that the neutron diffraction refinements for samples of $La_2NiMnO_6$ support the presence of $Ni^{2+}$ and $Mn^{4+}$,[9] whereas Bull and coworkers conclude that $Ni^{3+}$ and $Mn^{3+}$ are present.[10] Although they assign different Ni/Mn oxidation states, both of these neutron diffraction studies report the existence of high and low temperature $La_2NiMnO_6$ phases, and provide evidence for partial Ni/Mn ordering in the B-site sublattice. The refinements of these neutron diffraction studies also suggest the presence of anti-site defects (i.e., Ni and Mn atoms are not ordered perfectly) within the ordered sublattice, which where accounted for improving the refinement of the data.

With respect to crystal structure, $La_2NiMnO_6$ is biphasic with a high temperature rhombohedral phase that transforms at low temperatures to a monoclinic or orthorhombic phase, the latter phase depending on the arrangement of the Ni/Mn B-site sublattice.[2] A random distribution of Ni and Mn over the octahedral sites of the perovskite structure occurs for rhombohedral (*R-3c*) and orthorhombic (*Pbnm*), while an ordering of Ni and Mn into a distinguishable sites can be accommodated in rhombohedral (*R-3* or *R-3m*) and monoclinic (*P2$_1$/n*) space groups. Both the high temperature rhombohedral and low temperature monoclinic/ orthorhombic phases coexist over a wide temperature range, including room temperature.[3]

Synthetic conditions determine the atomic order of the B-site sublattice, and, as a result, the ferromagnetic properties of $La_2NiMnO_6$ samples.[2,11] Samples prepared at low temperatures by the glycine-nitrate method at temperatures less than 500 °C in air are orthorhombic and have a well-defined $T_C \approx 150$ K. Annealing the samples at 1300 °C produces a $La_2NiMnO_6$ phase with rhombohedral symmetry with a $T_C \approx 280$ K. Samples synthesized at intermediate temperatures (500-1300 °C) generates a mixture of the ferromagnetic phases. The low-$T_C$ phase of $La_2NiMnO_6$ has been attributed to trivalent oxidation states, $Ni^{3+}$ and $Mn^{3+}$, while the high-$T_C$ phase has been shown to be characteristic of atomically ordered $Ni^{2+}$ and $Mn^{4+}$.[2]

Epitaxial thin film samples of La$_2$NiMnO$_6$ (LNMO) have been deposited previously on different substrates by the pulsed laser deposition (PLD) method.[12,13] These thin film samples of La$_2$NiMnO$_6$ display ferromagnetic behavior with magnetic transitions temperatures similar to those observed for bulk samples (T$_C$ ~ 280 K). Variable temperature Raman spectroscopy indicates the presence of short-range ordering of *Ni/Mn* up to 400 K in these thin film samples.[14] Nonetheless, the structure of these films as deposited on cubic substrates, where the growth process and lattice-substrate mismatch can introduce epitaxial strain, sample inhomogeneities, and phase separation, has yet to be studied. In particular, the long-range *Ni/Mn* ordering on the B-site sublattice has yet to be characterized.

Here we report on the epitaxial growth of high-quality LNMO thin films and their structural and physical properties. In particular, transmission electron microscopy (TEM) was used to study the microstructure of the thin films samples. Multiple LNMO thin films were grown on (001)-orientated SrTiO$_3$ (STO) substrates. Stoichiometric La$_2$NiMnO$_6$ was employed as a target, which was synthesized by conventional solid state methods. The films were deposited 650 – 750 °C by the PLD technique using a KrF excimer laser (248 nm, 3 Hz) under a 100 mTorr atmosphere of flowing oxygen under dynamic vacuum. On average, 3000 pulses yielded films with a thickness of 70 nm. The crystalline structure of the thin film samples was examined by X-ray diffraction (XRD) using a Seifert 3000P diffractometer (Cu K$_\alpha$, λ = 1.5406 Å). A Philips X'Pert diffractometer was used for the in-plane XRD measurements of the film samples. Figure 1 shows a typical XRD *θ-2θ* pattern for a LNMO film grown on STO at 720 °C. The peaks were indexed based on a pseudo-cubic unit cell and only the peaks corresponding (0*k*0) reflections (where *k* = 1, 2, 3…) were observed, which indicates that the out-of-plane lattice parameters is a multiple of the perovskite subcell parameter (a$_{sub}$ ≈ 3.9 Å). The perovskite subcell lattice parameter was also confirmed by the electron microscopy study. The absence of diffraction peaks from secondary phases or randomly orientated grains indicates the preferential orientation of the films. The full-widths-at-half-maximum (FWHM) is 0.19 ° for the (020) reflection (inset of Figure 1), as measured by rocking-curve analysis. The in-plane orientation, as evaluated by the XRD Φ-scan of the LNMO (103) reflection of the cubic subcell, shows four peak separated by 90 °, indicating that the LNMO film is epitaxial with respect to the substrate.

Measurements of magnetization ($M$) versus applied magnetic field ($H$) and temperature ($T$) were performed on all samples using a superconducting quantum interference device magnetometer (SQUID). As shown in Figure 2, the in-plane loop recorded at 10 K shows a well-defined hysteresis with a coercive field approximately equal to 920 Oe and a saturation field close to 5 kOe, which represents a saturation magnetization of 5 $\mu_B$/formula unit. Moreover, the M(T) (insert of Figure 2) recorded with a 500 Oe applied magnetic field, shows a clear magnetic transition at 270 K and a minor one close to 150 K. Both magnetic transitions have been reported previously for different spin states of Ni and Mn. In particular, the high temperature transition has been reported for previous thin film samples of LNMO.[12,13] The DC-electrical properties of the film samples were measured by a physical property measurement system (PPMS) in four-probe configuration. As expected, the thin film samples were semi-conducting and are similar to those reported previously for bulk LNMO samples.[3]

In order to address the ordering of Ni/Mn in the B-site sublattice, TEM was used to study the thin film samples at room temperature. To prepare the samples for study, a film was scratched off of the substrate and deposited onto a holey carbon film supported by a copper grid. Energy dispersive spectroscopy (EDS) on the microscopes (JEOL 200CX, JEOL 2010F, and TOPCON 002B) was used to analyze the stoichiometry of the thin film samples. Analysis of numerous flakes confirmed the samples to have a cationic ratio of 2/1/1 for La/Mn/Ni, within the limit of accuracy of the EDS technique. Electron diffraction (ED) analysis was used to probe the microstructure of the thin film samples, and the ED patterns exhibit a set of intense reflections corresponding to that of the lattice parameter of the perovskite subcell. The superposition of the film and substrate implies that the subcell parameter ($a_{sub}$) is very close to 3.89 Å. Reconstruction of the reciprocal space was carried out by tilting around the $<100>^*_{sub}$ and $<110>^*_{sub}$ equivalent directions. A super-cell was evidenced with a ≈ c ≈ $a_{sub}\sqrt{2}$ and b ≈ $2a_{sub}$. The conditions limiting the reflection indicate a centered "$I$-type" cell with the conditions $0kl : (k + l = 2n)$ and, at 90 ° tilting around $\vec{b}^*$, $hk0 : h = 2n$ (Figure 3). Such conditions imply the absence of a four-fold axis, and are consistent with the space groups *Imma* and *Im2a* or *I2/a* (*c* unique axis with γ≈90 °), which commonly are observed for distorted perovskites. Based on the in-plane ED patterns, the "*I*-type" phase can be indexed either as a [010] zone axis in the case of *Imma* and *Im2a* (or *I2/a* with γ ≈ 90 °) or as a [101] zone axis in the case of *I2/a* ($a ≈ 5.50$ Å, $b ≈ 7.78$ Å, $c ≈ 5.45$ Å, and γ ≈ 89 °). The presence of the mirror plane *a* in these possible "*I*-type" unit cells is not consistent with the complete long-range ordering of the *Ni/Mn* B-site sublattice, and may provide evidence for anti-site defects, which are present in bulk samples.[9,10]

In a few areas of the film, estimated to be a few percent, ED patterns exhibit extra reflections. These reflections keep the same cells' dimensions a ≈ c ≈ $a_{sub}\sqrt{2}$ and b≈$2a_{sub}$ with α ≈ β ≈ γ ≈ 90 ° (within the limit of accuracy), but violate "*I*-centering" symmetry and involve a "*P*-type" space group. All variants of the distorted "*P*-type" cell are observed in the basal plane of the ED patterns. The close values of $a/\sqrt{2}$, $b/2$ and $c/\sqrt{2}$ favor the formation of twinning domains, which are difficult to detect in the "*I*-type" phase, and multiple orientations of these nano-domains can be observed in the "*P*-type" phase. As shown in Figure 4, high resolution images illustrate that this "*P*-type" phase is stabilized in the form of nano-sized domains spread out in the "*I*-type" matrix. While short-range *Ni/Mn* order cannot be ruled out within these nano-domains, the multiple orientations of the "*P*-type" phase nano-domains within the "*I*-type" matrix argue against a consistent long-range *Ni/Mn* order in these thin film samples. Such antiphase boundaries, however, can be aligned with a modest applied magnetic field leaving 360 ° spin rotation across an antiphase boundary.[2] It should be noted that the presence of both the majority "*I*-type" and minority "*P*-type" phase at room temperature confirms that, similar to bulk samples, thin film LNMO samples are biphasic. Indeed, preliminary TEM studies indicate that the minority "*P*-type" phase becomes more prevalent at low temperatures at the expense of the "*I*-type" phase. These results are consistent with reports of bulk La$_2$NiMnO$_6$ transitioning from a room temperature mixture of *P*2$_1$/*n* and *R-3* to pure *P*2$_1$/*n* at 3.5 K.[3] Further work is in progress to explore the temperature-dependence phase transition and its relation to physical properties.

In conclusion, epitaxial La$_2$NiMnO$_6$ thin films have been deposited on (001)-orientated STO using the pulsed laser deposition technique. The material is semiconducting and ferromagnetic with a Curie temperature close to 280 K, a coercive field of 920 Oe, and a saturation magnetization of 5 $\mu_B$/formula unit, all values which are similar to those reported for bulk and other thin film La$_2$NiMnO$_6$ samples. Transmission electron microscopy, conducted at room temperature, reveals a majority "*I*-centered" phase ($a \approx c \approx a_{sub}\sqrt{2}$ and $b \approx 2a_{sub}$) coexisting with domains of a minority "*P*-type" phase, which are dispersed in the "*I*-type" matrix. The presence of the mirror plane *a* in the "*I*-type" phase and the multiple orientation of the "*P*-type" nano-domains argues against complete long-range ordering of the *Ni/Mn* sublattice for these thin film samples.

This work is carried out in the frame of the NoE FAME (FP6-5001159-1), the STREP MaCoMuFi (NMP3-CT-2006-033221) and the STREP CoMePhS (NMP4-CT-2005-517039) supported by the European Community and by the CNRS, France. Partial support from the ANR (NT05-1-45177, NT05-3-41793) is also acknowledged.

...

**Figure captions:**

Figure 1: $\Theta$–$2\Theta$ XRD scan of a typical film. S indicated the reflections of the SrTiO$_3$ substrate. Insert shows a rocking-curve recorded around the 020 reflection of the film.

Figure 2: (M-H) loop recorded at 10K. Insert displays the (M-H) curve recorded with 500Oe. Magnetic field is applied parallel to the [100] direction of the substrate.

Figure 3: ED patterns obtained along (a): [100] and (b): [001] zone axes, respectively.

Figure 4: a) HREM image observed along one of the <100>$_{sub}$ direction. Most of the film exhibits a "I-type" matrix where, for such a direction, only the perovskite subcell is evidenced (see Fourier Transform in c). In this image, a nano-sized "P-type "domain exhibits a 7.6Å periodicity as attested by the Fourier Transform in b).The darker blotches are point defects.

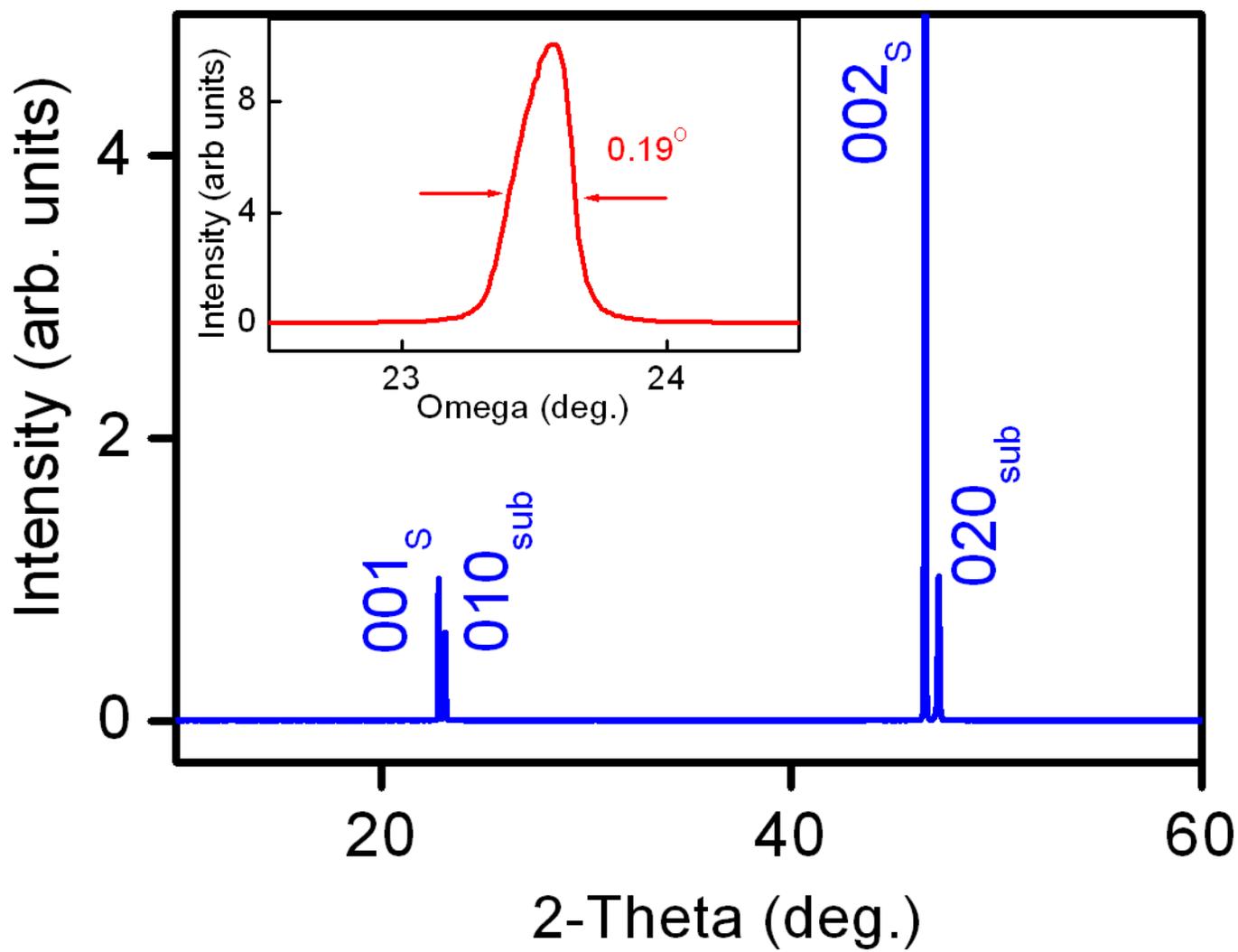

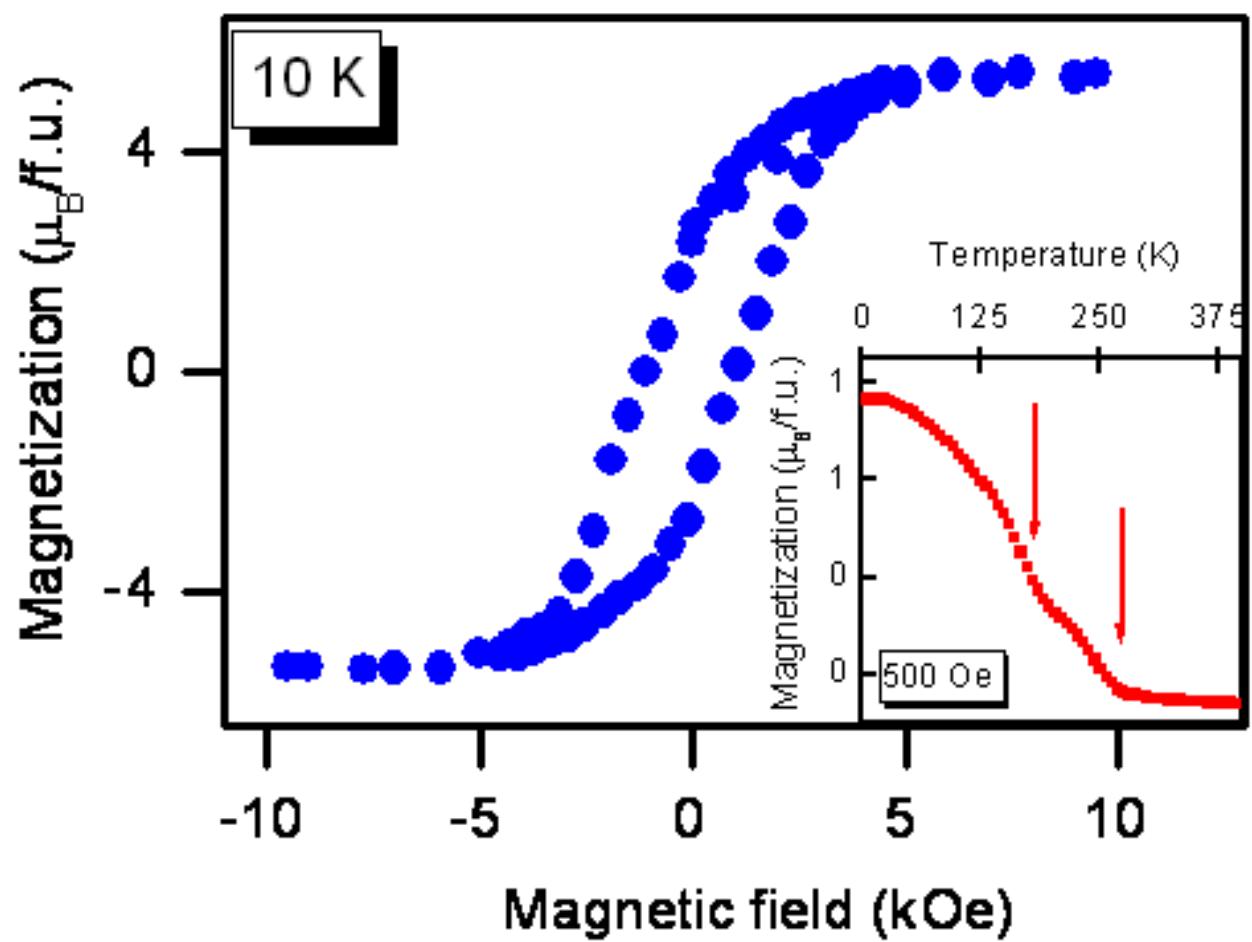

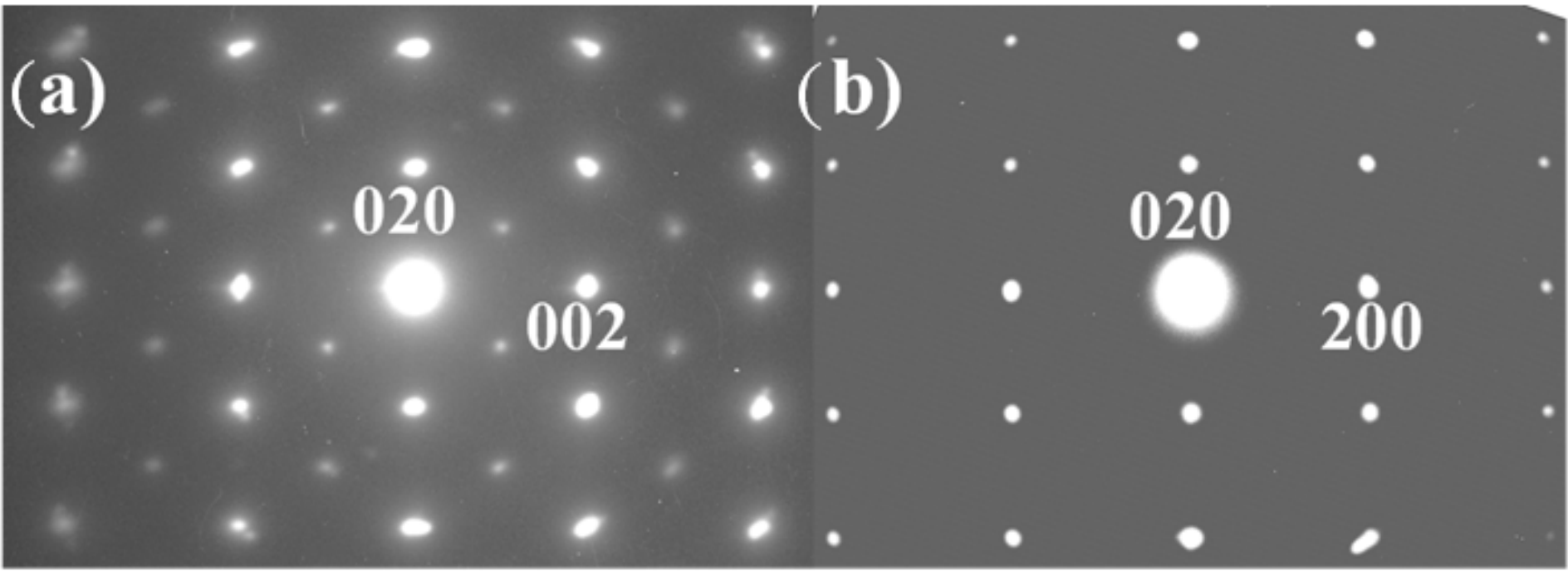

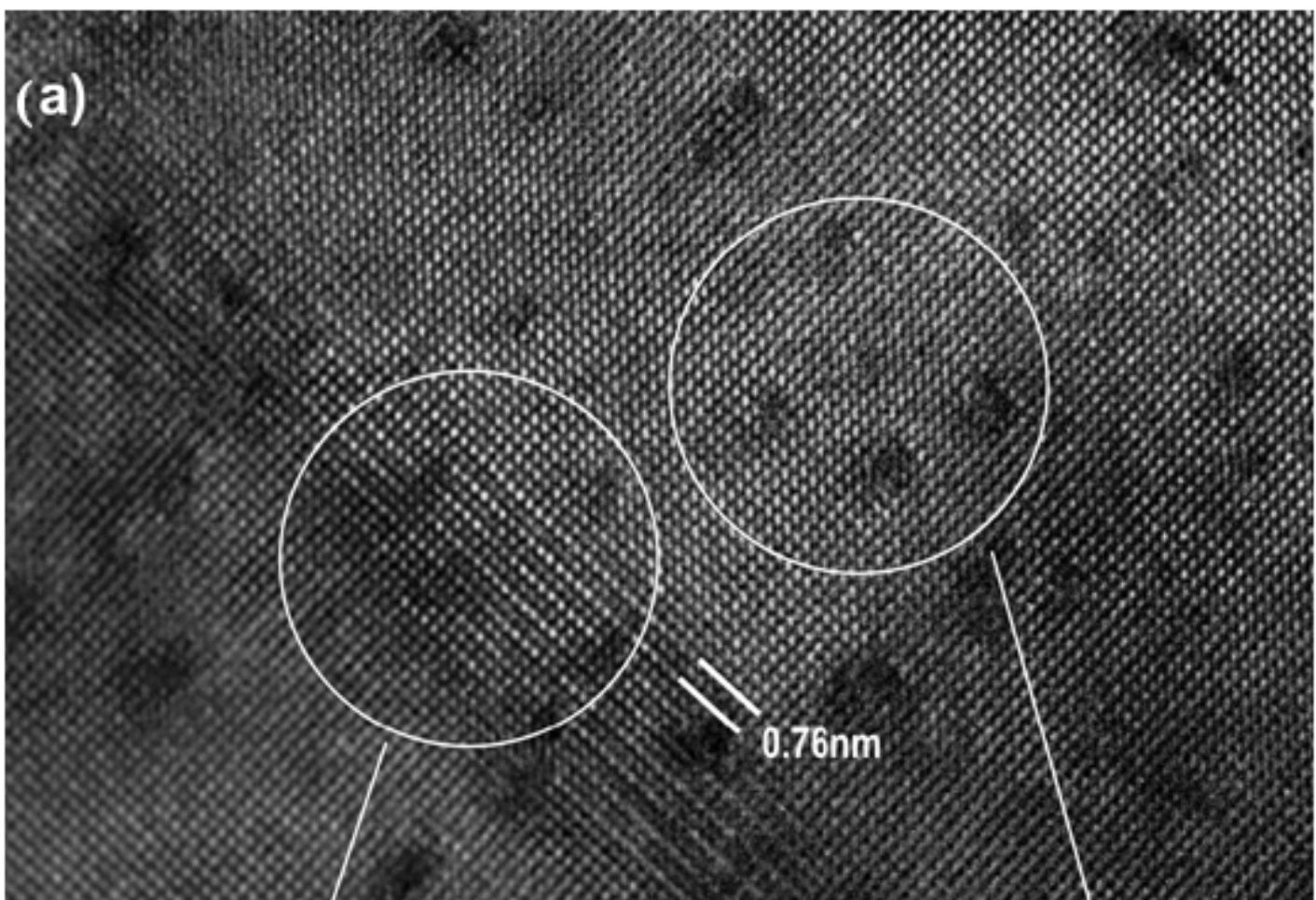
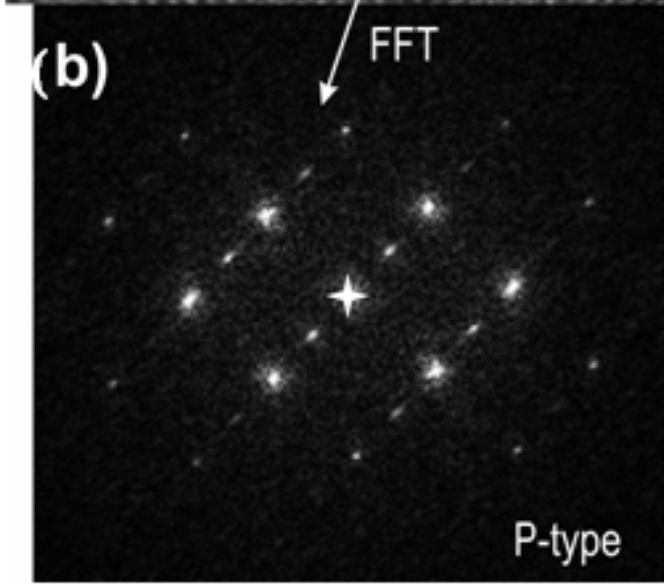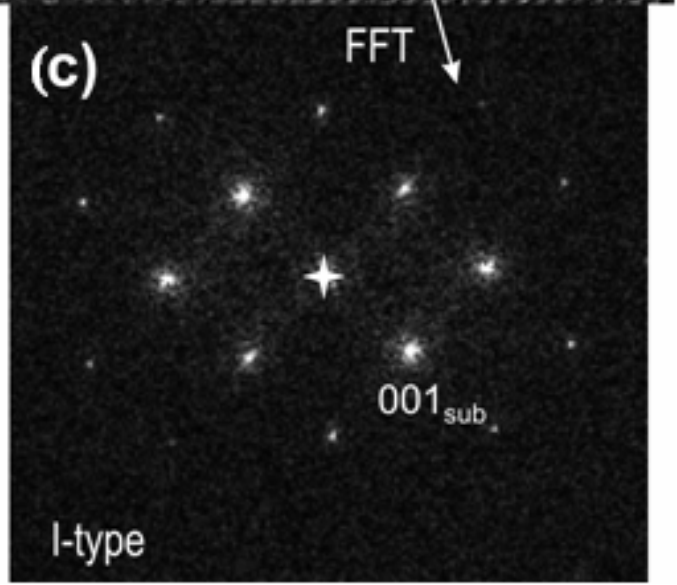